\documentclass[preprint,
superscriptaddress,
amsmath,amssymb,
aps,dvipsnames,usenames,prl]{revtex4-1}

\usepackage{graphicx}
\usepackage{dcolumn}
\usepackage{bm}
\usepackage[utf8]{inputenc}
\usepackage[T1]{fontenc}
\usepackage{mathptmx}
\usepackage{float}

\begin{document}

\title{Recent twists in twisted light:\\
A perspective on optical vortices from dielectric metasurfaces}

\author{Marco Piccardo*}
\affiliation{Center for Nano Science and Technology, Fondazione Istituto Italiano di Tecnologia, Milan 20133, Italy}

\author{Antonio Ambrosio*}
\affiliation{Center for Nano Science and Technology, Fondazione Istituto Italiano di Tecnologia, Milan 20133, Italy}

\collaboration{*Authors to whom correspondence should be addressed: marco.piccardo@iit.it, antonio.ambrosio@iit.it.}

\date{\today}

\begin{abstract}
Optical vortices are the electromagnetic analogue of fluid vortices studied in hydrodynamics. In both cases the traveling wavefront, either made of light or fluid, is twisted like a corkscrew around its propagation axis---an analogy that inspired also the first proposition of the concept of optical vortex. Even though vortices are one of the most fundamental topological excitations in nature, they are rarely found in their electromagnetic form in natural systems, for the exception of energetic sources in astronomy, such as pulsars, quasars and black holes. Mostly optical vortices are artificially created in the laboratory by a rich variety of approaches. Here we provide our perspective on a technology that shook-up optics in the last decade---metasurfaces, planar nanostructured metamaterials---with a specific focus on its use for molding and controlling optical vortices.
\end{abstract}

\maketitle

\begin{figure*}
    \centering
    \includegraphics[width=0.9\textwidth]{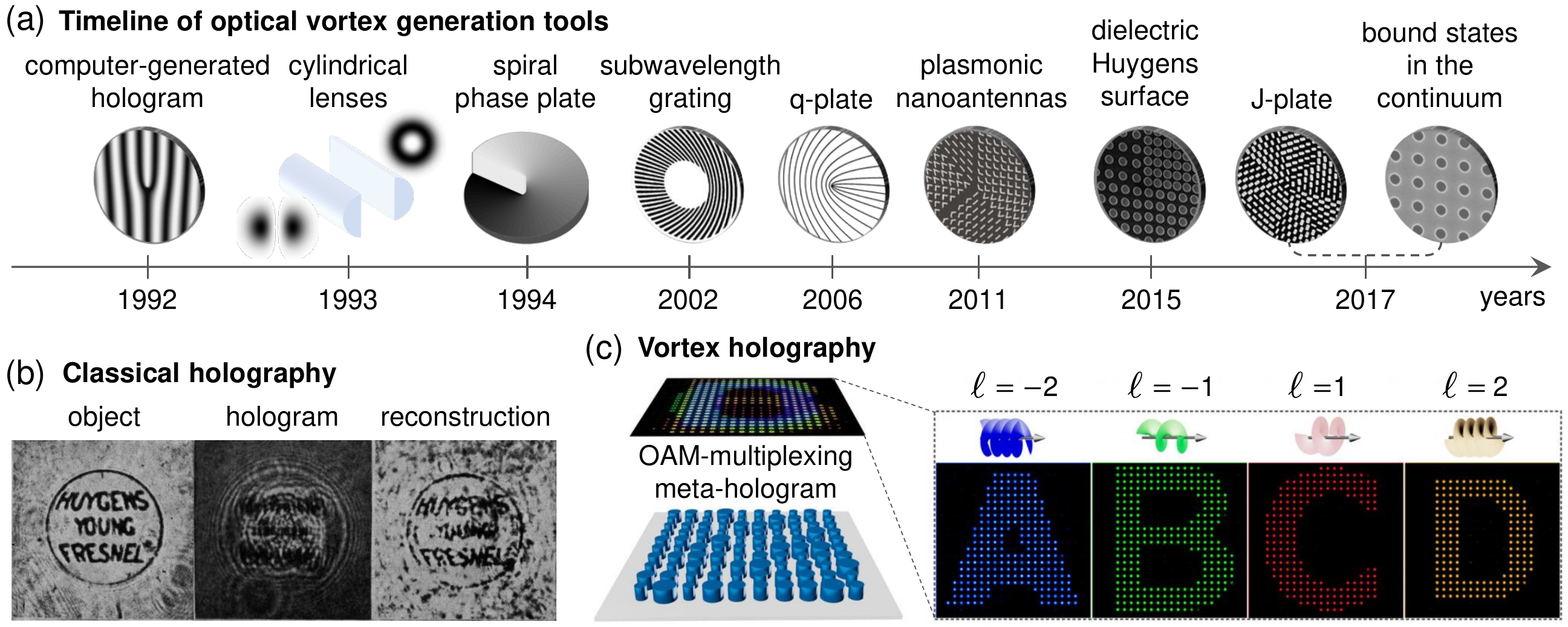}
    \caption{(a) Appearance in chronological order of different optical devices for the generation of optical vortices. The four most recent elements are based on metasurfaces. (b) First demonstration of classical holography by Dennis Gabor in 1948 showing the original object, its holographic plate and the reconstructed field. Image adapted from Ref. \citenum{GABOR1948}. (c) Metasurface-based orbital angular momentum holography using vortex beams as information carriers. The nanopillars of a GaN meta-hologram preserve the orbital angular momentum information allowing to reconstruct different images depending on the topological charge of the illuminating beam. Image adapted from Ref. \citenum{Ren2019} and reproduced under Creative Commons Attribution 4.0 International License.}
    \label{fig_timeline}
\end{figure*}

Since the introduction of the concept of optical vortex (OV) by Coullet \textit{et al.}\cite{Coullet1989} in 1989 a number of methods to generate such beams have emerged. In Fig. \ref{fig_timeline}a we present a timeline tracing the appearance of OV generation tools in the last 30 years. We restrain our analysis to optical components external to a laser cavity, which typically benefit from compactness and simplicity of implementation (for active sources of OVs see, e.g., Ref. \citenum{Forbes2017}). The first tool for OV generation was based on computer-generated holograms\cite{Heckenberg1992} (CGHs). These are physical holographic plates that, upon illumination by a reference plane-wave, produce a vortex beam. The interference pattern of the hologram may appear as a spiral or a pitch fork, depending on whether the propagation of the illuminating beam is on- or off-axis. Nowadays such approach is mainly implemented with spatial light modulators \cite{Forbes2016} (SLMs) rather than physical plates. OVs from CGHs appeared just before the pioneer work by Allen \textit{et al.} of 1992, which recognized that photons carry orbital angular momentum (OAM) and proposed an experiment to measure the OAM-induced mechanical torque exerted on a mode converter based on cylindrical lenses. Soon after, in 1993, such mode converter was experimentally realized for the generation of Laguerre-Gaussian (LG) beams from Hermite-Guassian (HG) inputs \cite{Beijersbergen1993}. A different tool based on spatially-variant phase accumulation \cite{Beijersbergen1994} appeared in 1994, known as spiral phase plate. This component owes its name to the characteristic spiral structure of the surface causing an azimuthally-dependent phase delay of the incident beam. Today the evolution of spiral phase plates is represented by spiral phase mirrors \cite{Fickler2016} that, thanks to the progress in direct surface machining, can implement extremely large values of azimuthal modulation, producing beams with OAM charges over 10,000. In the early 2000's new devices based on birefringent media exploiting the geometric (or Pancharatnam-Berry) phase were introduced: subwavelength gratings using dielectric grooves \cite{Biener2002}, and the $q$-plate based on liquid crystals \cite{Marrucci2006}. $Q$-plates have retardance elements of few $\mu$m in size thus are limited in resolution, but at the same time benefit from electrical tunability. With subwavelength gratings one can achieve much smaller feature sizes, however beam shaping is limited when the grooves are continuous as in Ref. \citenum{Biener2002}.

Subwavelength gratings \cite{Hasman2005} can be considered as one of the precursors of metasurfaces, which are planar nanostructured materials implementing a spatially-varying optical response that allows to mold the light wavefront in all its properties, i.e. amplitude, phase and polarization. A key concept in metasurfaces appeared in 2011 with the introduction by Yu $\textit{et al.}$ of anomalous reflection and refraction based on abrupt phase discontinuities \cite{Yu2011}. Interestingly, the first practical realization of this concept was used to generate an OV with an OAM charge of 1 in the mid-infrared. Such metasurface was based on metallic V-shaped nanoantennas, which suffer from severe ohmic loss at visible frequencies. Another example of plasmonic OAM metasurface but operating in the visible range was presented in Ref. \citenum{Karimi2014}. A great improvement in the performance of metasurfaces was enabled by switching to low-loss, high-refractive index dielectric nanostructures. In 2015 silicon-based Huygens' dielectric metasurfaces for OV generation in the near-infrared were introduced \cite{Chong2015}, followed in 2017 by another type of dielectric metasurfaces based on $q$-plates operating at visible wavelengths \cite{Devlin2017a}. Also in 2017 a completely different OV generation tool was demonstrated---the $J$-plate \cite{Devlin2017}---providing an unprecedented control over the total angular momentum ($J$) of an optical beam. The $J$-plate will be examined in more details in this Perspective. Thanks to their flexibility in terms of wavefront manipulation, dielectric metasurfaces hold great promise for the generation and modulation of OVs but a challenge still remains for the fabrication of large area, broadband devices. For reference, the largest area metalens devices fabricated to date, though not involving OV generation, have a typical size of few cm and are monochromatic \cite{She2018}. An alternative to metasurfaces for large area and broadband generation of OVs is represented by spin-to-OAM conversion in metallic \cite{Kobayashi2012a} and dielectric \cite{Radwell2016} cones, though this approach is limited to small values of OAM charge. We conclude this historical excursus on OV generation by mentioning a class of optical components that appeared more recently based on photonic bound states in continuum \cite{Kodigala2017,Wang2020} (BICs). These are photonic crystals---in some sense also a particular type of metasurface \cite{Huang2020}---with subwavelength structures arranged to control the electromagnetic resonances of the crystal slab, effectively turning it into a high-quality-factor resonator without the need to form a physical cavity. BICs are intrinsically connected with topological charges allowing to generate OVs without any real-space chiral structure \cite{bahari2017integrated}. Since they are very sensitive to symmetry breaking they can be easily controlled by tuning the symmetry of the pump beam. This has recently allowed to achieve ultrafast modulation (THz rate) of vortex beams in perovskite BIC metasurfaces \cite{Huang2020}.

The concept of metasurface is closely related to that of holography. Classical holography, invented in 1948 by Dennis Gabor \cite{GABOR1948}, allows to encode into a holographic plate both the intensity and phase information of a wave scattered by an object and, upon illumination, to reconstruct the original field (Fig. \ref{fig_timeline}b). Thanks to the method of CGHs the interference pattern may be also calculated for virtual objects, thus allowing to reconstruct fields even for objects that never existed. The progress of holography was favored by the widespread diffusion of spatial light modulators (SLMs) but remained limited in performance due to their typically large pixel size. With the advent of nanofabrication and the complete manipulation of light enabled by nanostructured materials, the new field of metasurface holography was born \cite{Huang2018}. However, until very recently the information carriers of light in holographic systems were limited to polarization, wavelength and time. In 2020 the concept of OAM holography was introduced \cite{Fang2020} and demonstrated with metasurfaces \cite{Ren2019}. The key idea is to utilize a spatial sampling of the digital hologram that is OAM-dependent, which allows to preserve the OAM information in the holographic image. This can be understood as the fields at the hologram plane and image plane constitute a Fourier pair, thus the product of the hologram with the OAM reconstruction beam occurring at the metasurface corresponds to a convolution between the holographic image and the Fourier transform of the OAM beam at the image plane. When the image is sampled with the correct OAM-dependent period matching the one of the illuminating beam this convolution becomes an OAM-pixelated version of the image, where each pixel corresponds to a vortex. Vortex holography not only allows to preserve the OAM information but also to achieve selectivity and multiplexing: by combining together multiple OAM-selective meta-holograms it is possible to reveal different holographic images from the same device using reconstruction beams with different OAM charges (Fig. \ref{fig_timeline}c), which holds promise for ultrahigh-capacity and holographic systems with high-security encryption.

\begin{figure*}
    \centering
    \includegraphics[width=0.95\textwidth]{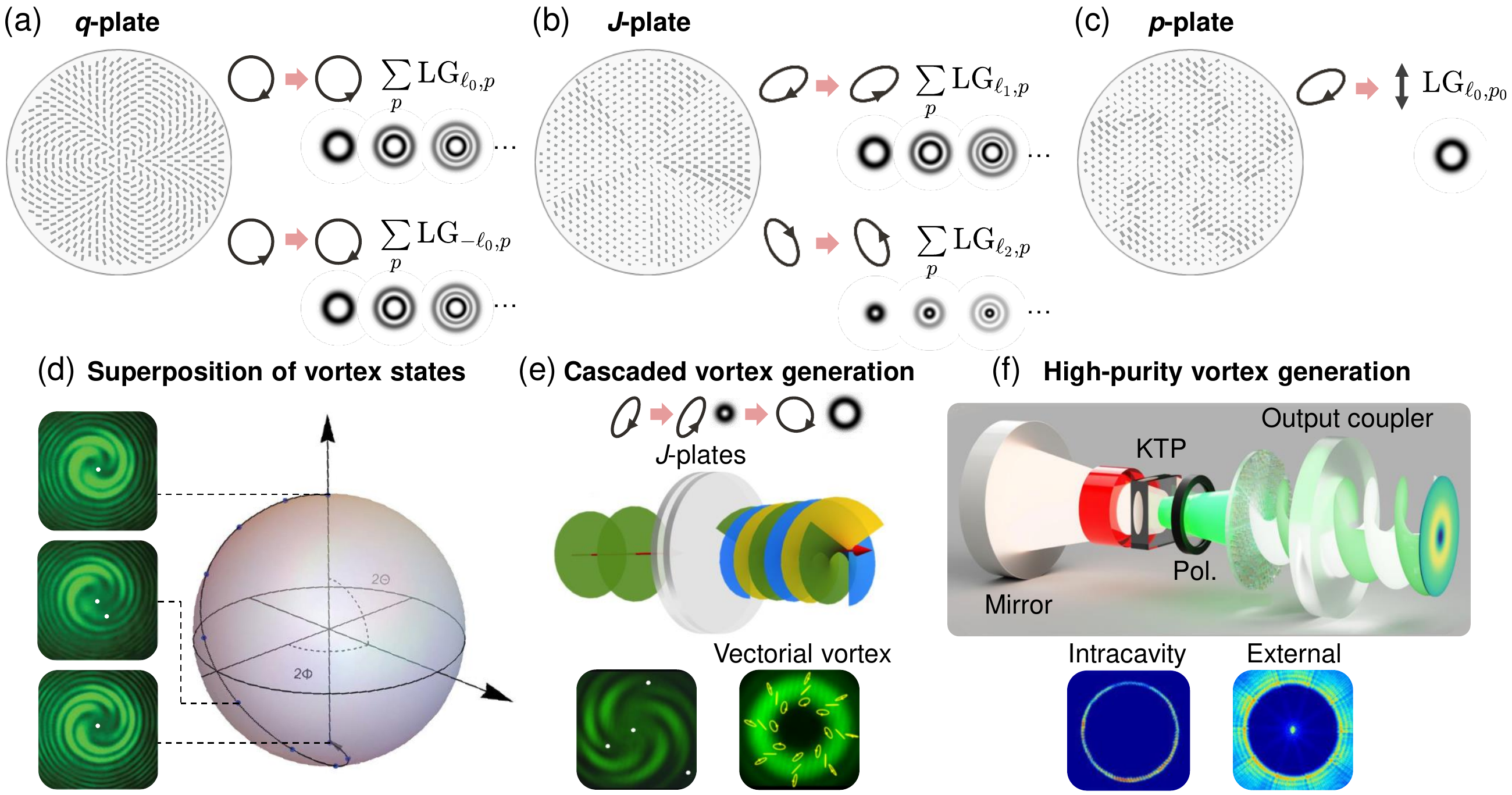}
    \caption{(a)-(c), Comparison of different dielectric metasurfaces for optical vortex generation: (a), a $q$-plate, generating from circular polarization states conjugate orbital angular momentum states, in this example with $|\ell_0|=5$; (b), a $J$-plate, which generalizes the spin-to-orbital angular momentum conversion allowing to generate two beams with independent values of topological charge, here $\ell_1 = +5$ and $\ell_2=+1$, from arbitrary orthogonal polarization states; (c), a $p$-plate, which gives full control over the radial ($p$) mode of vortex beams and generating in this example a mode with $\ell_0=+5$ and $p_0=0$. (d)-(f), Some key experimental demonstrations obtained using $J$-plates: (d), different superpositions of vortex states obtained from a single $J$-plate represented on the higher-order Poincaré sphere, with off-axis singularities marked as white dots on the spiral interference patterns; (e) cascaded generation of optical vortices by pairing two $J$-plates, giving origin to superpositions of total angular momentum states with multiple off-axis singularities (bottom-left) or non-separable vector vortex beams with spatially-varying polarization (bottom-right); (f) generation of optical vortices with high purity by the embeddement of a $J$-plate in a cavity for second-harmonic generation, showing in particular the difference in beam purity for vortex generation inside and outside the cavity (bottom). Images in (d), (e) and (f) adapted/reprinted from Ref. \citenum{Devlin2017}, \citenum{Huang2019} and \citenum{Sroor2020}, respectively.}
    \label{fig_qJp}
\end{figure*}

Metasurfaces for OV generation have progressively evolved over the last decade in terms of complexity and functionality. In Fig. \ref{fig_qJp}a-c we compare three different types of dielectric devices based on birefringent (form-birefringence) nanopillars---the $q$-plate, the $J$-plate and the $p$-plate---that can be distinguished based on the polarization states on which they operate and on the properties of the OAM beams they generate. The $q$-plate \cite{Devlin2017a} is limited to the circular polarization (CP) basis and a single absolute value of OAM charge (Fig. \ref{fig_qJp}a): it converts right- (left-) CP into left- (right-) CP with OAM given by $\ell_0$ ($-\ell_0$). The $J$-plate represents a considerable generalization of the $q$-plate \cite{Devlin2017}: it converts an arbitrary pair of orthogonal polarization states into their conjugate states, which have flipped handedness, with two independent values of OAM charges $\ell_1$ and $\ell_2$ (Fig. \ref{fig_qJp}b). Both the $q$-plate and the $J$-plate are phase-only metasurfaces, i.e. they mold only the phase of the polarization-converted output beam but its amplitude distribution remains identical to that of the input beam. In other words when a $q$-plate or $J$-plate is illuminated with a Gaussian beam, the output in the near field, i.e. right after passing the metasurface, will be a Gaussian with an azimuthal phase profile. Since this is not an eigenmode of free space nature will correct for the missing amplitude modulation by creating a superposition of radial ($p$) modes upon propagation to the far field \cite{Sephton:16}. Thus these OV generators typically produce impure vortex modes consisting of several concentric rings of intensity (Fig. \ref{fig_qJp}a,b), which limits their detection efficiency in both classical and quantum OAM applications \cite{Nape2020}. Recently another type of dielectric metasurface allowing to overcome this issue was introduced. Called $p$-plate, as it allows to control the radial mode of an OAM beam, it converts an arbitrary polarization state into two orthogonal linearly-polarized state, one of which contains a desired combination of $\ell_0$ and $p_0$ values of a LG beam \cite{piccardo2020arbitrary} (Fig. \ref{fig_qJp}c). The linear polarization state containing the pure vortex mode can be simply selected by means of a linear polarizer following the metasurface.

Further insights into the different principles of these three devices can be gained by considering their representation in Jones calculus. The matrix corresponding to a birefringent nanopillar of a dielectric metasurface is
\begin{equation}
    M = \mathrm{e}^{i \psi} \left( \begin{matrix} 
                1 & 0 \\
                0  & \mathrm{e}^{i  \Delta\phi}
                \end{matrix} \right)
\label{eq_M}
\end{equation}
where $\psi$ is the global accumulated phase and $\Delta\phi$ is the phase retardance among the ordinary and extraordinary polarization components of the input beam. In addition, a pillar can be rotated by an angle $\alpha$ with respect to the frame of reference of the input beam, thus the pillar effect is described by the operator $R(-\alpha) M R(\alpha)$, where $R(\alpha)$ is the rotation matrix that changes frame of reference from that of the input beam to the pillar's one. In the case of a $q$-plate each nanopillar acts as a half-wave plate converting CP, thus $\Delta\phi=\pi$. The azimuthal phase profile is embedded in the pillars rotation angle thanks to the geometrical phase as e$^{\pm i 2\alpha}$ where $\alpha=q\varphi + \alpha_0$, $\varphi$ being the azimuthal coordinate of the device, and $\alpha_0$ and $q$ being constants \cite{Marrucci2006,Devlin2017a}; for instance, if the pillars complete 2.5 rotations over a circular loop around the center of the device then the OAM charge imparted by the $q$-plate is $|\ell_0|=5$ (Fig. \ref{fig_qJp}a). The global phase $\psi$, unused in $q$-plates, is exploited in a $J$-plate giving much more flexibility in the spin-to-orbital angular momentum conversion. The expressions of the birefringence matrix (Eq. \ref{eq_M}) and pillar angle for a $J$-plate in the case of arbitrary input polarizations is not compact \cite{Devlin2017} but in the simple case of linear input polarizations we have $\psi=\ell_1 \varphi$, $\psi + \Delta\phi=\ell_2 \varphi$ and $\alpha=0$, which clearly shows the fundamental link between the pillar birefringence and the two independent values of OAM charges imparted by the $J$-plate to orthogonal polarizations. As it can be noticed from Fig. \ref{fig_qJp}a,b the nanopillars do not vary in dimensions or orientation angle along the radial lines of $q$-plates and $J$-plates, meaning that the output beams are not structured along the radial directions, which is the origin of the modal impurity of the vortex beams generated by these devices. In contrast, a $p$-plate employs different pillars along a radial direction (Fig. \ref{fig_qJp}c) to implement the amplitude transmission mask corresponding to a desired radial mode. Also in this case the general expression of the nanopillar parameters is rather complicated \cite{piccardo2020arbitrary}, but considering a linear input polarization state \cite{Divitt2019} we have that the amplitude of the transmitted beam through an orthogonal polarizer is $A = \mathrm{sin}(2\alpha)$, its phase is e$^{i\psi}$, while the phase retardance needs to be set to $\Delta\phi=\pi$, as in a half-waveplate.

Since $q$-plates implemented with dielectric metasurfaces are rather limited in functionality and $p$-plates were introduced only very recently, we will concentrate in the following on some key demonstrations of OV control based on $J$-plates. One of the most attractive aspects of $J$-plates is the possibility to create superpositions of arbitrary pairs of OAM states, which is of interest in both quantum and classical optics experiments \cite{Forbes2019entangled}. Given a linear combination of orthogonal polarization states $\left| \lambda^+\right>$ and $\left| \lambda^-\right>$ at its input, a $J$-plate generates the superposition of states $\alpha \mathrm{e}^{i \ell_1 \varphi} \left| \lambda^+\right> + \beta \mathrm{e}^{i \ell_2 \varphi} \left| \lambda^-\right>$, where $\alpha$ and $\beta$ are the linear combination coefficients. In doing so the $J$-plate maps the Poincaré sphere onto the higher-order Poincaré sphere \cite{Milione2011} (HOPS), which represents both spin and orbital angular momentum states. As an example, Fig. \ref{fig_qJp}d shows a HOPS having at the north pole a state with right-CP and $\ell_1=+3$, and at the south pole a state with left-CP and $\ell_2=+4$. The corresponding spiral interference patterns show a number of spiral arms equal to the topological charge of the states. By varying the linear combination of input CPs, the output state from the $J$-plate follows a trajectory on the HOPS. The superposition of vortex states is characterized by the appearance of an off-axis singularity, which shifts progressively to the beam center as the south pole of the HOPS is approached \cite{Devlin2017} (Fig. \ref{fig_qJp}d).

A $J$-plate is also capable of mapping a HOPS onto another HOPS. By cascading multiple $J$-plates it is thus possible to convert at each stage both the pair of orthogonal polarization states and the OAM charge (Fig. \ref{fig_qJp}e), giving an unprecedented control over total angular momentum (TAM) states. For instance, by combining two $J$-plates one can obtain a superposition of four different TAM states with several off-axis singularities \cite{Huang2019}. OAM modes can also be paired with cascaded $J$-plates in a non-separable manner, so to form vector vortex beams \cite{Ndagano2018,Bao2020} with spatially-variant polarization (Fig. \ref{fig_qJp}e). Such output states cannot be written as a single direct product of spin and orbital angular momentum states: the measurement of one of these states affects the other, e.g. turning a polarizer changes the spatial pattern of the transmitted beam, making them an interesting demonstration of the concept of classical entanglement of light \cite{Forbes2019entangled}. 

As mentioned above, a single $J$-plate allows to control only the azimuthal properties of an optical beam (Fig. \ref{fig_qJp}b) and thus generates impure vortex beams consisting of multiple radial modes (Fig. \ref{fig_qJp}f, bottom-right panel). However this problem was overcome in a recent experiment \cite{Sroor2020} showing that the inclusion of a $J$-plate in a second-harmonic generation cavity with a KTiOPO$_4$ (KTP) crystal pumped by a Nd:YAG laser permits to generate pure vortex beams, i.e. with radial mode $p=0$ and arbitrary OAM charge (Fig. \ref{fig_qJp}f, bottom-left panel). This demonstrates a coherent source of powerful, high-purity OAM modes, which can also be produced in a superposition of states as a different form of chiral light not observed before from lasers.

As we have seen, there has been a great deal of momentum recently in OAM research. It is our belief that, thanks to the technological progress in nanofabrication, the orbit around the many interesting concepts recurring in this field will soon spiral down to land on the long-sought, real-world OAM applications \cite{Rubinsztein-Dunlop2016}. Metasurfaces are definitely a key technology to help succeed in this endeavor, thanks to their full control of the properties of light. Now that the development of single metasurfaces for optical vortices generation has reached satisfactory results, more effort should be dedicated to the integration of these plates in compound systems, in the spirit of Moiré metasurfaces \cite{Ohno2018} or combining, for instance, metasurfaces with tunable optical plates. Finally, we see as particularly promising the following branches of OAM research as powered by the metasurface technology: quantum optics, thanks to the improvements in purity, efficiency and power of the OV sources \cite{Sroor2020,piccardo2020arbitrary}; optical tweezing, with light structured not only in the spatial but also temporal domain \cite{omatsu2017wavelength,Rego2019}; and optical communications, benefitting from the recent advances in multiplexing, cryptography \cite{Fang2020,Ren2019} and ultrafast vortex modulation \cite{Huang2020}.

\begin{acknowledgments}
This work has been financially supported by the European Research Council (ERC) under the European Union’s Horizon 2020 research and innovation programme ``METAmorphoses'', grant agreement no. 817794.
\end{acknowledgments}

\section*{Data Availability}
The data that support the findings of this study are available from the corresponding author upon reasonable request.

\bibliography{vortices_perspective_bib}

\end{document}